\documentclass[preprint,prc,aps,showpacs,showkeys,groupedaddress,floatfix]{revtex4}
\usepackage{epsfig}
\usepackage{dcolumn}
\usepackage{bm}% bold math
\usepackage{graphics}
\usepackage{graphicx}
\usepackage{amssymb}
\usepackage{amsmath}
\begin{document}
\title{The Energy Eigenvalues of the Two Dimensional Hydrogen Atom in a Magnetic Field}
\author{A.~Soylu$^{1,2}$}
\email{E-mail:asimsoylu@gmail.com}
\author{O.~Bayrak$^{1,3}$}
\author{I.~Boztosun$^{1}$}
\affiliation{$^{1}$ Department of Physics, Faculty of Arts and
Sciences, Erciyes University, Kayseri, Turkey \\ $^{2}$ Department
of Physics, Faculty of Arts and Sciences, Nigde University, Nigde,
Turkey \\ $^{3}$ Faculty of Arts and Sciences, Department of
Physics, Bozok University, Yozgat, Turkey}
\begin{abstract}
In this paper, the energy eigenvalues of the two dimensional
hydrogen atom are presented for the arbitrary Larmor frequencies by
using the asymptotic iteration method. We first show the energy
eigenvalues for the no magnetic field case analytically, and then we
obtain the energy eigenvalues for the strong and weak magnetic field
cases within an iterative approach for $n=2-10$ and $m=0-1$ states
for several different arbitrary Larmor frequencies. The effect of
the magnetic field on the energy eigenvalues is determined
precisely. The results are in excellent agreement with the findings
of the other methods and our method works for the cases where the
others fail.
\end{abstract}
 \maketitle
\section{Introduction}
The study of the two dimensional hydrogen atom in a magnetic field
has been a subject of considerable interest over the years. Within
the framework of the non-relativistic quantum mechanics, many works
have been carried out in order to solve the eigenvalue equation and
to find out the correction on the energy eigenvalues in the presence
of a constant magnetic field  \cite{victor1,v1,v2,v3,v4}. The
solution of this problem is very interesting and popular because of
the technological advances in nanofabrication technology which has
enabled the creation of low-dimensional structures such as quantum
wires, quantum dots and quantum wells in semiconductor physics.
Recent developments in nanostructure technology has also permitted
one to study the behavior of electrons and impurities in quasi
two-dimensional configurations (quantum wells)
\cite{victor1,v1,v2,v3,v4,ref2,ref3,ref4,ref5,ref6,ref7}.

The canonical Hamiltonian for a charge moving in a constant magnetic
field can be written as:
\begin{equation}\label{ham1}
H = \frac{1}{2\mu}(\textbf{p}+ \frac{e}{c}\textbf{A})^2  + V(r)
\end{equation}
where $\mu$ is the mass, $e$ is the electric charge, $\textbf{p}$ is
the momentum of the particle, \textbf{A} is the vector potential,
$c$ is the light velocity and V(r) is the cylindrical potential
\cite{fluge}. The hamiltonian for the 2D Hydrogen atom in the
magnetic field includes the Coulomb interaction $-Z/r$ between a
conduction electron and donor impurity center when a constant
\textbf{B} magnetic field is applied perpendicular to the plane of
the motion. If the vector potential in the symmetric gauge is chosen
as $\textbf{A}= \frac{1}{2}\textbf{B} \times \textbf{r}$, the full
Hamiltonian for this system can be derived, in the CGS system and in
atomic units $\hbar=\mu=e=1$, as:

\begin{equation}\label{ham2}
H = \frac{1}{2}( - i\nabla   + \frac{1}{2} \textbf{B}  \times
\textbf{r} )^2 -\frac{Z}{r}
\end{equation}
and the Schr\"{o}dinger equation becomes

\begin{equation}\label{ham3}
H\varphi  = \frac{1}{2}( - i\nabla   + \frac{1}{2}\textbf{B} \times
\textbf{ r} )^2 \varphi  - \frac{Z}{r}\varphi  = i\partial _t
\varphi  = E\varphi
\end{equation}
Since the problem pertains to  two dimensions, it is adequate to
study in polar coordinates $(r,\phi)$ within the plane and to use
the following ansatz for the eigenfunction
\begin{equation}\label{ef}
    \varphi(r)=\frac{e^{im\phi}}{\sqrt{2\pi}}\frac{R(r)}{\sqrt{r}},
    \quad m=0, \pm1, \pm2...
\end{equation}
Here, the radial wavefunction $R(r)$ must satisfy the following
radial Schr\"{o}dinger equation:
\begin{equation}\label{sch}
    \frac{d^{2}R(r)}{dr^{2}}+2\left[(E-m\omega_{L})-\frac{(m^{2}-\frac{1}{4})}{2r^{2}}
    -\frac{1}{2}\omega_{L}^{2}r^{2}+\frac{Z}{r}\right]R(r)=0
\end{equation}
where $\omega_{L}=B/2c$ is the Larmor frequency, $E$ is the energy
eigenvalue and $m$ is the eigenvalue of the angular momentum.

As it is seen from equation (\ref{sch}), we need an effective
potential of $\alpha r^{-2}+\beta r^{-1}+\gamma r^{2}$ type, which
is a hybrid of Coulomb plus harmonic oscillator potential, in order
to describe the two dimensional hydrogen atom in a magnetic field.
This potential can not be solved analytically except for particular
cases and there are no general closed form solutions to equation
(\ref{sch}) in terms of the special functions \cite{bagrov}. There
are analytic expressions for the eigenvalues for particular values
of $w_{L}$ and $m$ \cite{Taut1,Taut}.

Therefore, in this paper, in order to find the energy eigenvalues
for the two dimensional hydrogen atom in a constant magnetic field
with the arbitrary Larmor frequencies $w_L$
\cite{MacD,mustafa,mustafa2,victor2},  we use a more practical and
systematic method, called the Asymptotic Iteration Method (AIM) for
different $n$ and $m$ quantum numbers. This is precisely the aim of
this paper.

In the next section, we briefly outline AIM with all necessary
formulae to perform our calculations. In section \ref{apply}, we
first apply AIM to solve the Schr\"{o}dinger equation for the case
$\omega_{L}=0$: no magnetic field and to obtain an analytical
expression for any $n$ and $m$ states. Then, we show how to solve
the resulting Schr\"{o}dinger equation for the case $w_{L}\neq0$:
strong and weak magnetic fields where there are no analytical
solutions. Here, for any $n$ and $m$ quantum numbers, we show the
effect of the magnetic field on the energy eigenvalues and compare
our results with the findings of other methods \cite{Taut}. Finally,
section \ref{conclude} is devoted to our summary and conclusion.
\section{Basic Equations of the Asymptotic Iteration Method (AIM)}
\label{AIM} AIM is proposed to solve the second-order differential
equations of the form \cite{hakan,orhan,mesut}.
\begin{equation}\label{diff}
  y''=\lambda_{0}(x)y'+s_{0}(x)y
\end{equation}
where $\lambda_{0}(x)\neq 0$  and the functions, s$_{0}$(x) and
$\lambda_{0}$(x), are sufficiently differentiable. The differential
equation (\ref{diff}) has a general solution \cite{hakan}
\begin{equation}\label{generalsolution}
  y(x)=exp\left(-\int^{x}\alpha(x^{'})dx^{'}\right )\left [C_{2}+C_{1}
  \int^{x}exp\left(\int^{x^{'}}(\lambda_{0}(x^{''})+2\alpha(x^{''}))dx^{''}\right )dx^{'}\right]
\end{equation}
if $k>0$, for sufficiently large $k$, we obtain the $\alpha(x)$
values from
\begin{equation}\label{quant}
\frac{s_{k}(x)}{\lambda_{k}(x)}=\frac{s_{k-1}(x)}{\lambda_{k-1}(x)}=\alpha(x),
\quad k=1,2,3,\ldots
\end{equation}
where
\begin{eqnarray}\label{iter}
\lambda_{k}(x) & = &
\lambda_{k-1}'(x)+s_{k-1}(x)+\lambda_{0}(x)\lambda_{k-1}(x) \quad
\nonumber \\
s_{k}(x) & = & s_{k-1}'(x)+s_{0}(x)\lambda_{k-1}(x), \quad \quad
\quad \quad k=1,2,3,\ldots
\end{eqnarray}
The energy eigenvalues are obtained from the quantization condition.
The quantization condition of the method together with equation
(\ref{iter}) can also be written as follows
\begin{equation}\label{quantization}
  \delta_{k}(x)=\lambda_{k}(x)s_{k-1}(x)-\lambda_{k-1}(x)s_{k}(x)=0, \quad \quad
k=1,2,3,\ldots
\end{equation}

For a given potential, the radial Schr\"{o}dinger equation is
converted to the form of equation (\ref{diff}). Then, s$_{0}(x)$ and
$\lambda_{0}(x)$ are determined and s$_{k}(x)$ and $\lambda_{k}(x)$
parameters are calculated by using equation (\ref{iter}). The energy
eigenvalues are determined by the quantization condition given by
equation (\ref{quantization}).

\section{AIM Solution for the Two Dimensional Hydrogen Atom}
\label{apply} Applying the scale transformation $r=r_{0}\rho \quad
 (r_{0}=\frac{1}{2Z})$ to equation (\ref{sch}) and by using the following
\emph{ansatz}:

\begin{equation}\label{ansatz}
    \varepsilon=\frac{(E-m\omega_{L})}{2Z^{2}}, \quad
    \beta^{2}=\frac{\omega_{L}^{2}}{16Z^{4}} , \quad \quad  l'(l'+1)=m^{2}-\frac{1}{4}
    \end{equation}
we can easily find
\begin{equation}\label{schaim}
\frac{d^{2}R(\rho)}{d\rho^{2}}+\left[\varepsilon+\frac{1}{\rho}-\beta^{2}\rho^{2}-\frac{l'(l'+1)}{\rho^{2}}\right]R(\rho)=0
\end{equation}
In what follows, we show how to obtain the energy eigenvalues from
this equation for two different cases, depending on the values of
$\omega_{L}$ and show the effect of the $\omega_{L}$ on the
eigenvalues.

\subsection{Case $\omega_{L}=0$ : no magnetic field}
When $\omega_{L}=0$, equation (\ref{schaim}) becomes
\begin{equation}\label{schaim1}
\frac{d^{2}R(\rho)}{d\rho^{2}}+\left[-\varepsilon'^2+\frac{1}{\rho}-\frac{l'(l'+1)}{\rho^{2}}\right]R(\rho)=0
\end{equation}
where $-\varepsilon'^2=\frac{(E-m\omega_{L})}{2Z^{2}}$. In order to
solve this equation with AIM, we should transform this equation to
the form of equation (\ref{diff}). Therefore, the reasonable
physical wave function we propose is as follows
\begin{equation}
R(\rho ) = \rho ^{l' + 1} e^{ - \varepsilon' \rho } f(\rho )
\end{equation}
equating it into equation (\ref{schaim1}) leads to
\begin{equation}
\frac{{d^2 f(\rho )}}{{d\rho ^2 }} = \left(\frac{{2\varepsilon'\rho
- 2l' - 2}}{\rho }\right)\frac{{df(\rho )}}{{d\rho }} +
\left(\frac{{2\varepsilon' l'+ 2\varepsilon'  - 1}}{\rho
}\right)f(\rho )
\end{equation}
where $ \lambda _0=2(\frac{{\varepsilon' \rho  - l' - 1}}{\rho })$
and $ \ s_0 =\frac{{2\varepsilon 'l' + 2\varepsilon' - 1}}{\rho }$.
By means of equation(\ref{iter}), we may calculate $\lambda _k (\rho
)$ and $ \ s_k (\rho)$. This gives
\begin{eqnarray}\label{lambda}
\lambda_0&=&2\left(\frac{{\varepsilon' \rho- l'- 1}}{\rho }\right) \nonumber \\
s_0&=& \frac{{2\varepsilon 'l' + 2\varepsilon'  - 1}}{\rho} \quad \nonumber \\
\lambda_1&=&\frac{{10l'+6-6\rho\varepsilon'l'-6\varepsilon'\rho-\rho+4\varepsilon'^2\rho^2+4l'^2}}{\rho^2}
\nonumber \\
s_1&=&\frac{{-10\varepsilon'l'-6\varepsilon'+3+4\varepsilon'^2l'\rho-4\varepsilon'l'^2+4\varepsilon'^2\rho-2\varepsilon'\rho+2l'}}{\rho^2}
\\ \nonumber
\quad\quad\quad\quad...etc
\end{eqnarray}

Combining these results with the quantization condition given by
equation (\ref{quantization}) yields
\begin{eqnarray}
 \frac{s_0 }{\lambda _0 } & = & \frac{s_1 }{\lambda _1 }\,\,\,\,\,\, \Rightarrow
\,\,\,\,\,\,(\varepsilon')_{0}=\frac{1}{2(l'+1)}            \nonumber \\
 \frac{s_1 }{\lambda _1 }  & = & \frac{s_2 }{\lambda _2 }\,\,\,\,\,\, \Rightarrow
\,\,\,\,\,\, (\varepsilon')_{1}=\frac{1}{2(l'+2)}           \nonumber \\
 \frac{s_2 }{\lambda _2 }  & = & \frac{s_3 }{\lambda _3 }\,\,\,\,\,\, \Rightarrow
\,\,\,\,\,\,(\varepsilon')_{2}=\frac{1}{2(l'+3)} \\
\ldots \emph{etc} \nonumber
\end{eqnarray}
If the above expressions are generalized, $\varepsilon'$ turns out
as
\begin{equation}\label{quant4}
(\varepsilon')_{n}=\frac{1}{2(l'+n+1)}\quad\quad\quad\quad\quad
n=0,1,2,3...
\end{equation}
If one inserts values of $\varepsilon'$ and $l'$ into equation
(\ref{quant4}), the eigenvalues of the $2D$ hydrogen atom in the
case $w_{L}=0$ becomes
\begin{equation}\label{ew0}
E_{n}=-\frac{1}{2(|m|+n+\frac{1}{2})^2}
\end{equation}
This analytical formula is in agreement with the previous works
\cite{fluge}. We discuss the results of the $w_{L}=0$ case together
with the findings of the case $w_{L}\neq0$ in the next subsection.
\subsection{Case $w_{L}\neq0$ : strong and weak magnetic fields}
Before applying AIM to this problem, we have to obtain asymptotic
wavefunction and then transform equation (\ref{schaim}) to an
amenable form for AIM. We transform equation (\ref{schaim}) to
another Schr\"{o}dinger form by changing the variable as
$\rho=u^{2}$ and then by inserting $R(u)=u^{1/2}\chi(u)$ into the
transformed equation. Thus, we get another Schr\"{o}dinger form
which is more suitable for an AIM solution:
\begin{equation}\label{insert}
\frac{d^{2}\chi(u)}{du^{2}}+ \left[4\varepsilon
u^{2}+4-4\beta^{2}u^{6}-\frac{\Lambda(\Lambda+1)}{u^{2}}\right]\chi(u)=0
\end{equation}
where $\Lambda=2l'+\frac{1}{2}$. It is clear that when $u$ goes to
zero, $\chi (u)$ behaves like $u^{\Lambda+1}$ and $\exp \left(
-\frac{\alpha }{4}u^{4}\right)$ at infinity, therefore, the
wavefunction for this problem can be written as follows:

\begin{equation}
\chi(u)=u^{\Lambda+1}\exp \left( -\frac{\alpha }{4}u^{4}\right) f(u)
\end{equation}%
If this wave function is inserted into equation (\ref{insert}), we
have the second-order homogeneous linear differential equation in
the following form
\begin{equation}
f^{\prime \prime }=2\left( \alpha u^{3}-\frac{\Lambda+1}{u}\right)
f^{\prime }+\left( \left[ \left( 2\Lambda+5\right) \alpha -4\varepsilon %
\right] u^{2}-4\right) f
\end{equation}%
Where $\alpha =2\beta $. By comparing this equation with equation
(\ref{diff}), $\lambda_{0}(u)$ and $s_{0}(u)$ values can be written
as below
\begin{equation}\label{lambdas}
\lambda _{0}=2\left( \alpha u^{3}-\frac{\Lambda+1}{u}\right), \quad
\quad s_{0}=\left[ \left( 2\Lambda+5\right) \alpha -4\varepsilon
\right] u^{2}-4
\end{equation}

In AIM, we calculate the energy eigenvalues from the quantization
condition given by equation (\ref{quantization}). It is important to
point out that the problem is called ``exactly solvable" if this
equation is solvable at every $u$ point. In our case, since the
problem is not exactly solvable, we have to choose a suitable
$u_{0}$ point and to solve the equation $\delta
_{k}(u_{0},\varepsilon )=0 $ to find $\varepsilon $ values. In this
work, we obtain the $u_{0}$ from the maximum point of the asymptotic
wavefunction which is the same as the root of $\lambda _{0}(u)=0$,
thus $u_{0}=\left( \frac{\Lambda+1}{\alpha }\right) ^{1/4}$. The
results obtained by using AIM are shown in Tables \ref{table1} and
\ref{table2} in comparison with the results of Ref. \cite{Taut} for
$Z=1$, $n=2-10$, $m=0$ and $m=1$ with different Larmor frequencies
$\omega_{L}$. Ref. \cite{Taut} was able to solve this equation
analytically for particular values of $\omega_{L}$, $n$ and $m$
quantum numbers. However, he could not obtain the ground state
energy eigenvalue and the energy eigenvalues also diverged for
$\omega_{L}$=0: No solution could be obtained. In Table
\ref{table3}, we have shown the eigenvalues for several Larmor
frequencies for the ground state and second excited states, which
could not be obtained by Ref. \cite{Taut}. In order to show that our
method can obtain energy eigenvalues for arbitrary Larmor
frequencies, we have calculated the energy eigenvalues for a few
arbitrary Larmor frequencies with $m=0$, $m=1$ and $n=1-3$ values in
Tables \ref{table4} and \ref{table5}. The first lines of Tables
\ref{table4} and \ref{table5} show the results for the case
$w_{L}=0$ with different quantum numbers $n$ and $m$. We present
only $m=0$, $m=1$ and $n=1$ to $3$ for illustration, but any value
of $n$ and $m$ can be obtained.

\section{Conclusion}\label{conclude}
In this paper, we have shown an alternative method to obtain the
energy eigenvalues for the two dimensional hydrogen atom without the
magnetic field and in a constant magnetic field for arbitrary Larmor
frequencies for various $n$ and $m$ quantum numbers within the
framework of the asymptotic iteration method. We have first applied
AIM to solve the radial Schr\"{o}dinger equation for the case
$\omega_{L}=0$: no magnetic field and have obtained an analytical
expression for the energy eigenvalues for any $n$ and $m$ states.
Then, we have shown how to solve the resulting radial
Schr\"{o}dinger equation for the case $w_{L}\neq0$: strong and weak
magnetic fields. This equation cannot be solved analytically,
therefore, we have shown how to obtain the energy eigenvalues by an
iterative approach within the  framework of AIM. By comparing these
two cases ($w_{L}=0$ and $w_{L}\neq0$), we have shown the effect of
the magnetic field on the energy eigenvalues in the tables and have
compared our results with the findings of other methods.

Besides showing the applicability of a new method to solve the
radial Schr\"{o}dinger equation in the magnetic field for any $n$
and $m$ quantum numbers, one of the novelties of this paper is that
we have shown that it is possible to obtain the ground state energy
eigenvalues where others works such as \cite{Taut} have failed to
obtain. We have also shown that it is possible to solve the
$w_{L}=0$ and $w_{L}\neq0$ case simultaneously where,  in general,
the $w_{L}=0$ case makes the energy eigenvalues diverge and
non-physical results are obtained (see Ref. \cite{Taut} for
details.)

It is clearly shown in this paper that the method presented in this
study is a systematic one and it is very efficient and practical to
obtain the eigenvalues for the Schr\"{o}dinger type equations in a
magnetic field and without the magnetic field. It is worth extending
this method to the solution of other problems.
\section*{Acknowledgments}
This work is supported by the Scientific and Technical Research
Council of Turkey (TÜBÝTAK) under the project number TBAG-2398 and
Erciyes University (FBT-04-16). Authors would like to also thank
Professor A. Ahmadov for stimulating discussions and useful comments
on the manuscript.

\newpage
\begin{table}
\begin{center}
\begin{tabular}{c}        %          \hline
 $m=0$                  \\
\begin{tabular}{|c|c|c|c|c}                  \hline
$n$ & $\omega_{L}^{-1} $ & $E$\cite{Taut} & $ E _{AIM}$     \\
\hline
 2  &  0.50000       & 4.0000000     &  4.0000000               \\ \hline
 3  &  3.00000       & 1.0000000     &  1.0000000               \\ \hline
 4  &  9.27200       & 0.4314060     &  0.4314064               \\ \cline{2-4}
    &  0.727998      & 5.4945200     &  5.4945207               \\ \hline
 5  &  21.1168       & 0.2367780     &  0.2367785               \\  \cline{2-4}
    &  3.88316       & 1.2876100     &  1.2876109               \\ \hline
 6  &  40.3133       & 0.1488340     &  0.1488343               \\ \cline{2-4}
    &  11.2570       & 0.5330000     &  0.5330020               \\ \cline{2-4}
    &  0.929632      & 6.4541700     &  6.4541668               \\ \hline
 7  &  68.6380       & 0.1019840     &  0.1019840               \\ \cline{2-4}
    &  24.6751       & 0.2836870     &  0.2836867               \\ \cline{2-4}
    &  4.68692       & 1.4935200     &  1.4935182               \\ \hline
 8  &  107.868       & 0.0741648     &  0.0741647               \\ \cline{2-4}
    &  45.9214       & 0.1742110     &  0.1742107               \\ \cline{2-4}
    &  13.0953       & 0.6109080     &  0.6109059               \\ \cline{2-4}
    &  1.11539       & 7.1723900     &  7.1723786               \\ \hline
 9  &  159.781       & 0.0563272     &  0.0563271               \\ \cline{2-4}
    &  76.7724       & 0.1172300     &  0.1172296               \\ \cline{2-4}
    &  28.0095       & 0.3213200     &  0.3213195               \\ \cline{2-4}
    &  5.43732       & 1.6552300     &  1.6552272               \\ \hline
 10 &  226.154       & 0.0442176     &  0.0442177               \\ \cline{2-4}
    &  119.005       & 0.0840301     &  0.0840301              \\ \cline{2-4}
    &  51.2233       & 0.1952240     &  0.1952236               \\ \cline{2-4}
    &  14.8274       & 0.6744290     &  0.6744268               \\ \cline{2-4}
    &  1.29016       & 7.7509600     &  7.7509774             \\ \hline
\end{tabular}
\end{tabular}
\end{center}
\caption{The energy eigenvalues obtained by using AIM for all
allowed Larmor frequencies $\omega_{L}$ used by Ref. \cite{Taut} for
$Z = 1$,  $n = 2-10$ and $m =0$.}\label{table1}
\end{table}
\begin{table}
\begin{center}
\begin{tabular}{c}        %          \hline
 $m=1$                  \\
\begin{tabular}{|c|c|c|c|c}                  \hline
$n$ & $\omega_{L}^{-1} $ & $E$\cite{Taut} &
$ E _{AIM}$     \\ \hline
 2  & 1.50000     & 2.6666700    &  2.6666667      \\ \hline
 3  & 7.00000     & 0.7142860   &  0.7142857      \\ \hline
 4  & 18.1394     & 0.3307720   &  0.3307717      \\ \cline{2-4}
    & 1.86059     & 3.2247800    &  3.2247835      \\ \hline
 5  & 36.6810     & 0.1909910   &  0.1907883      \\  \cline{2-4}
    & 8.34903     & 0.8384210   &  0.8384207      \\ \hline
 6  & 64.2985     & 0.1244200   &  0.1244197      \\ \cline{2-4}
    & 21.0161     & 0.3806600   &  0.3806606      \\ \cline{2-4}
    & 2.18539     & 3.6606800    &  3.6606737      \\ \hline
 7  & 102.855     & 0.0875018  &  0.0875018      \\ \cline{2-4}
    & 41.5559     & 0.2165760   &  0.2165757      \\ \cline{2-4}
    & 9.58910     & 0.9385660   &  0.9385656      \\ \hline
 8  & 154.096     & 0.0648944  &  0.0648947      \\ \cline{2-4}
    & 71.7176     & 0.1394360   &  0.1394358      \\ \cline{2-4}
    & 23.6998     & 0.4219450   &  0.4219444      \\ \cline{2-4}
    & 2.48615     & 4.0222800    &  4.0222838      \\ \hline
 9  & 219.800     & 0.0500456  &  0.0500455      \\ \cline{2-4}
    & 113.269     & 0.0971138  &  0.0971140      \\ \cline{2-4}
    & 46.1803     & 0.2381970   &  0.2381968      \\ \cline{2-4}
    & 10.7509     & 1.0231700    &  1.0231700      \\ \hline
 10 & 301.742     & 0.0397691  &  0.0397691      \\ \cline{2-4}
    & 167.984     & 0.0714353  &  0.0714354      \\ \cline{2-4}
    & 78.7673     & 0.1523470   &  0.1523475      \\ \cline{2-4}
    & 26.2373     & 0.4573650   &  0.4573641      \\ \cline{2-4}
    & 2.76930     & 4.3332300    &  4.3332248     \\ \hline
\end{tabular}
\end{tabular}
\end{center}
\caption{The same as table \ref{table1}, but for $m
=1$.}\label{table2}
\end{table}
\begin{table}
\begin{center}
\begin{tabular}{|c|c|c|c|c|} \hline
$\omega_{L}^{-1} $& $m$ &$n$ &  $E$\cite{Taut} &$ E _{AIM}$     \\
\hline 0.50000   &  0   &  1 &   -          & -1.459586     \\
\cline{3-5}
          &      &  2 & 4.00000      &  4.000000     \\ \cline{3-5}
          &      &  3 &   -          &  8.344348     \\ \hline
3.00000   &  0   &  1 &   -          & -1.979604     \\ \cline{3-5}
          &      &  2 &   -          &  0.180700     \\  \cline{3-5}
          &      &  3 & 1.00000      &  1.000000     \\  \hline
1.50000   &  1   &  1 &   -          &  1.200118     \\  \cline{3-5}
          &      &  2 & 2.66667      &  2.666666     \\ \cline{3-5}
          &      &  3 &   -          &  4.070242     \\ \hline
7.00000   &  1   &  1 &   -          &  0.002497     \\ \cline{3-5}
          &      &  2 &   -          &  0.387686     \\ \cline{3-5}
          &      &  3 & 0.714286     &  0.714285     \\ \hline
\end{tabular}
\end{center}
\caption{Several Larmor frequencies $\omega_{L}$ and corresponding
eigenvalues $E$ in $Z = 1$, $n = 2-10$ and $m =0-1$ and where $n=1$
is the ground state}.\label{table3}
\end{table}
\begin{table}
\begin{center}
\begin{tabular}{c}        %          \hline
 $m=0$                  \\
\begin{tabular}{|c|c|c|c|c}                  \hline
 $w_{L}$ & $n=1$     & $n=2$       & $n=3$ \\ \hline
  0 & -2.000000 & -0.222222   & -0.08000   \\ \hline
  0.050 & -1.999530 & -0.205286   & 0.001634   \\ \hline
  0.125 & -1.997078 & -0.135464   & 0.227218  \\ \hline
  0.214 & -1.991490 & -0.015458   & 0.541870  \\ \hline
  0.333 & -1.979644 &  0.180112   & 0.998682  \\ \hline
  0.500 & -1.955159 &  0.494679   & 1.676971   \\ \hline
  0.750 & -1.903352 &  1.017056   & 2.737039   \\ \hline
  1.166 & -1.784478 &  1.962502   & 4.564444  \\ \hline
  2.000 & -1.459586 &  4.000000   & 8.344352  \\ \hline
  4.500 & -0.121100 & 10.538252   & 20.029864 \\ \hline
\end{tabular}
\end{tabular}
\end{center}
\caption{Corresponding eigenvalues for $w_{L}=0$: no magnetic field
and for the arbitrary Larmor frequencies for the values of $m=0$ and
$n=1-3$, where n=1 is ground state, $n=2$ is the first excited
state, $n=3$ is the second excited state.}\label{table4}
\end{table}

\begin{table}
\begin{center}
\begin{tabular}{c}        %          \hline
 $m=1$                  \\
\begin{tabular}{|c|c|c|c|c}                  \hline
$w_{L}$ & $n=1$      & $n=2$    & $n=3$      \\ \hline
  0 & -0.222222  & -0.080000 & -0.040816   \\ \hline
  0.050 & -0.159232  & 0.043578 & 0.174858   \\ \hline
  0.125 & -0.031456  & 0.317382 & 0.60668   \\ \hline
  0.214 &  0.146176  & 0.677668 & 1.152122  \\ \hline
  0.333 &  0.406240  & 1.183896 & 1.903978  \\ \hline
  0.500 &  0.795208  & 1.918102 & 2.980934   \\ \hline
  0.750 &  1.406898  & 3.044950 & 4.61856   \\ \hline
  1.166 &  2.467582  & 4.959271 & 7.379467  \\ \hline
  2.000 &  4.675218  & 8.870088 & 12.981222  \\ \hline
  4.500 & 11.550624  &20.821926 & 29.981774   \\ \hline
\end{tabular}
\end{tabular}
\end{center}
\caption{The same as table \ref{table4}, but for
$m=1$.}\label{table5}
\end{table}

\bigskip

 \end{document}